\documentclass[prb,preprint]{revtex4-1}
\usepackage{graphicx}
\usepackage{amssymb}
\usepackage{amsmath}
\usepackage{scalerel}
\usepackage{subfigure}
\DeclareRobustCommand*{\citen}[1]{%
  \begingroup
    \romannumeral-`\x % remove space at the beginning of \setcitestyle
    \setcitestyle{numbers}%
    \cite{#1}%
  \endgroup
}

\newlength{\figwidth}
\newcommand{\fref}[1]{Fig.\,\ref{#1}}
\newcommand{\tref}[1]{Table\,\ref{#1}}
\newcommand{\eref}[1]{Eq.\,(\ref{#1})}

\newcommand{\sref}[1]{Sec.\!~\ref{#1}}

\newcommand{\cref}[1]{Ref.\,\citen{#1}}

\newcommand{\eg}{{\it e.g.}\ }
\newcommand{\etc}{{\it etc.}\ }
\newcommand{\etal}{{\it et al.}\ }

\newcommand{\aposteriori}{{\it a posteriori} }
\newcommand{\abinitio}{{\it ab initio} }
\usepackage{color}
\newcommand{\bs}{\mathsf{b}}

\newcommand{\eb}{\mathbf{e}}

\newcommand{\dm}{{\mathrm{d}}}

\newcommand{\Sigmab}{{\boldsymbol{\Sigma}}}
\newcommand{\rhob}{{\boldsymbol{\rho}}}
\newcommand{\kappab}{{\boldsymbol{\kappa}}}
\newcommand{\virial}{{\boldsymbol{\nu}}}

\newcommand{\heatfluxb}{{\mathbf{J}}}

\newcommand{\heatfluxv}{{\mathbf{J}}}

\newcommand{\conductivityb}{{\boldsymbol{\kappa}}}

\newcommand{\As}{\mathsf{A}}
\newcommand{\Bs}{\mathsf{B}}

\newcommand{\cb}{\mathbf{c}}

\newcommand{\fb}{\mathbf{f}}

\newcommand{\vb}{\mathbf{v}}

\newcommand{\xb}{\mathbf{x}}

\newcommand{\nb}{\mathbf{n}}
\newcommand{\Ib}{\mathbf{I}}

\newcommand{\Jb}{\mathbf{J}}

\newcommand{\cs}{\mathsf{c}}

\newcommand{\Nc}{\mathcal{N}}

\newcommand{\var}{\operatorname{var}}

\newcommand{\tr}{\operatorname{tr}}

\newcommand{\coef}{\mathsf{c}}
\newcommand{\data}{\mathsf{D}}
\newcommand{\basis}{\mathsf{b}}
\newcommand{\xdata}{\mathsf{r}}
\newcommand{\ydata}{\mathsf{k}}
\newcommand{\probability}{\mathit{p}}
\newcommand{\posterior}{\probability(\coef | \ydata, \xdata)} 
\newcommand{\likelihood}{\probability(\ydata | \xdata, \coef)}
\newcommand{\prior}{ \probability (\coef)}
\newcommand{\evidence}{\probability(\ydata | \xdata)}
\newcommand{\covar}{\operatorname{covar}}
\newcommand{\smallplus}{{\scaleobj{0.5}{+}}}
\newcommand{\smallminus}{{\scaleobj{0.5}{-}}}
\newcommand{\miller}[1]{$\langle${{#1}}$\rangle$}
\newcommand{\millersurface}[1]{\{{{#1}}\}}

\begin{document}
\title{\bf The influence of defects on the thermal conductivity of compressed LiF}
\author{R.E. Jones }
\affiliation{\it Sandia National Laboratories, P.O. Box 969, Livermore, CA 94551, USA}
\email{Corresponding author: rjones@sandia.gov}
\author{D.K. Ward}
\affiliation{\it Sandia National Laboratories, P.O. Box 969, Livermore, CA 94551, USA}

\begin{abstract}
Defect formation in LiF, which is used as an observation window in ramp and shock experiments, has significant effects on its transmission properties.
Given the extreme conditions of the experiments it is hard to measure the change in transmission directly.
Using molecular dynamics, we estimate the change in conductivity as a function of the concentration of likely point and extended defects using a Green-Kubo technique with careful treatment of size-effects. 
With this data, we form a model of the mean behavior and its estimated error;
then, we use this model to predict the conductivity of a large sample of defective LiF resulting from a direct simulation of ramp compression as a demonstration of the accuracy of its predictions.
Given estimates of defect densities in a LiF window used in an experiment, the model can be used to correct the observations of thermal energy through the window.
In addition, the methodology we develop is extensible to modelling, with quantified uncertainty, the effects of a variety of defects on thermal conductivity of solid materials.
\end{abstract}

\maketitle

\section{Introduction}

LiF is a ionic solid that is extremely transparent to short wavelength radiation due to its large band gap and, hence, is commonly used in observation optics for ramp and shock loading experiments.
These high-rate experiments\cite{crowhurst2011invariance} can reach strain rates of 10$^{10}$ s$^{-1}$ and inevitably introduce defects in the LiF windows which affect their transmission properties and, thus, the accuracy of the observations.
Given the extreme conditions of the experiments it is difficult to measure the changed in transmission directly during the loading process.
Since the experiments are non-equilibrium processes, it is likely that they result in a population of defects in excess of equilibrium concentrations derived from energies of formation and possibly of different types than observed at ambient conditions \cite{johnston1960dislocation,gupta1975dislocation}.

The myriad variety of possible lattice defects fall into the broad categories of: point (\eg vacancies), line (\eg dislocations), surface (\eg stacking faults) and volume (\eg voids) defects. 
Given the technological importance of LiF, it has been studied for some time.
In alkali halides, like LiF, the on-site Schottky vacancy is considered the predominant pre-existing defect\cite{catlow1979contribution,KittelCh18}.
Others types, such as Frenkel point defects\cite{schwartz1996electronic}, dislocation lines\cite{gilman1958dislocations,sproull1959effect}, areal stacking faults\cite{mohamed1974method,eyring1976structural} and crystal boundaries \cite{thacher1967effect}, have been observed, particularly after the material has undergone irradiation \cite{schwartz1996electronic}, plastic deformation \cite{andreev1968formation,verrall1977deformation}, or shock \cite{asay1972effects}.
Also there is evidence of strong interactions between defect types.
Specifically, ``decorated'' dislocations have been observed where the compressive part of the stress field of the dislocations attract point defects to dislocation cores \cite{bullough1970kinetics,sigle1988determination}.

Classical theory of the effect of defects on phonon scattering and thermal conductivity is typically limited to the linear influence with density expected at the dilute/well-separated limit.
This classical theory is more informative about size, mass, and frequency effects than defect correlation effects that are likely to occur at high defect densities.
It is typically based on solutions to the related elasticity problem. 
Characteristic scattering times and related properties in the context of kinetic theory have been derived for:
point impurities \cite{klemens1955scattering,abeles1963lattice,klemens1985theory,ratsifaritana1987scattering},
dislocations \cite{klemens1958thermal,ackerman1971lattice,ackerman1971phonon,madarasz1981phonon},
stacking faults \cite{klemens1957some}, and 
grain boundaries \cite{holland1963analysis}.
Summaries of these findings can be found in the texts by Ziman\cite{zimanCh8}, Srivastava\cite{srivastavaCh6} and Kaviany\cite{kavianyCh4}.
There is experimental evidence of significant departures from classical predictions in LiF, particularly in phonon-dislocation interaction \cite{anderson1972interaction,uzuki1972effect,roth1979interaction}.
There have also been computational efforts to extend or refine these theories, notably the work of Volz and co-workers on the thermal transport parallel to screw dislocations \cite{ni2014thermal}.
Their work touched on the fact that extended defects have geometric attributes beyond simple densities, such as line direction and Burger's vector for dislocations \cite{frank1957dislocations,HirthCh14}.
The Kapitza resistance of tilt boundaries and the like has also attracted study with molecular dynamics\cite{zheng2014phonon,goel2015kapitza,goel2016thermal}.

In this work, we focus on estimating the effect of defects that are likely appear in LiF under extreme loading have on thermal conductivity.
Given that these conditions are not easy to investigate experimentally, the types of defects and their densities and distributions are not well-known.
Since it is not feasible to survey all likely defects in this computational study, we chose those most likely to occur given the experimental evidence and are also amenable to simulation.
Using a Green-Kubo (GK) method and molecular dynamics (MD), we predict the dependence of thermal conductivity on Schottky vacancies, dislocations and tilt boundaries. 
In the process we apply the GK technique to computational cells much larger than typical in order to span the range from very high defect densities to the dilute limit and capture coordination effects. 
Using uncertainty quantification (UQ) techniques we build a model of this thermal conductivity data and compare its predictions to a direct application of GK to a large, crushed LiF system representative of the extreme experimental conditions of interest.

\section{Theory} \label{sec:theory}

As mentioned, our objective is to develop a model of the dependence of thermal conductivity on defects in the material.
A spatial distribution of defects can be represented \cite{hausdorff1921summationsmethoden} by its moments $\rho^{(n)}$
For example, a population of point defects with locations $\{\xb_i\}$ can be characterized by:
\begin{equation}
\rho^{(n)} = \frac{1}{V} \int \xb^n \sum_i \delta(\xb-\xb_i) \, \mathrm{d}^3x
       = \frac{1}{V} \sum_i \xb_i^n  
\end{equation}
where $\xb^0 = 1$, $\xb^1 = \xb$, $\xb^2 = \xb \otimes \xb$, \etc.
This formalism can be extended to lines $\xb_i(\xi)$, where $\xi$ is an arc-length coordinate, and surfaces $\xb_i(\xi_1,\xi_2)$, where $\xi_1,\xi_2$ are surface coordinates.
For instance, the 0-th moment, $\rho^{(0)}$, is the point density (number of point defects per volume) for point defects such as vacancies, the line density (length per volume) for line defects such as dislocations, or the areal density (area per volume) for interface defects such as grain boundaries.
The higher moments which characterize the relative distances between the defects and moments of mixed types, such as point and line defects, can also be constructed.

Hence, the dependence of the thermal conductivity tensor on defects can be expressed as $\kappab = \kappab(\rhob)$, where $\rhob = \{ \rho_a^{(0)}, \rho_a^{(1)}, \ldots, \rho_b^{(0)}, \ldots \}$ is compact notation for the moments of all the relevant defects.
If we expand the functional dependence of the (log of the) thermal conductivity tensor $\kappab$ on defects in moments of defect distribution, we obtain the series representation:
\begin{equation} \label{eq:model}
\log\left( \kappab_0^{-1} \kappab(\rhob) \right) =  \cb_0 + \sum_a \cb_a \rho_a + \sum_{a,b} \cb_{ab} \rho_a \rho_b + \ldots
\end{equation}
where the coefficients $\cb_a, \cb_{ab}$ are the sensitivities of $\log\left( \kappab_0^{-1} \kappab(\rhob) \right)$ to first and second order effects.
With the expectation that $\kappab \to \kappab_0$, the perfect lattice conductivity, as $\rhob \to \mathbf{0}$, and $\kappab \to \mathbf{0}$  as $\| \rhob \|  \to \infty $, we see that $\Ib = \exp \cb_0$ and we expect that every element of $\cb_a$ is negative.
Also we expect that the components of the conductivity tensor $\kappab$ are diagonal.
Hidden in this notation is the usual thermodynamic dependence on pressure/density and temperature.
These variables will be held fixed for simplicity and specified in the Methods section.

Using MD and GK we can only obtain thermal conductivity data at specific densities with uncertainty,
This uncertainty arises from finite sampling of two ensembles: the thermodynamic ensemble and the defect distribution at fixed $\rhob$.
Given the typical MD cell sizes, the densities that are feasible to represent are typically larger than the defect densities and, hence, data from the strongly correlated regime is used in predictions of thermal conductivity in the dilute regime.
For these reasons we are motivated to create a model with inherent uncertainty quantification and error estimation.

Following Kennedy and O'Hagan's seminal paper\cite{kennedy2001bayesian}, Rasmussen and Williams' text \cite{rasmussen2006gaussianCh2} and our previous work \cite{rizzi2013uncertaintyI,rizzi2013uncertaintyII}, we use Bayesian regression to calibrate the coefficients of \eref{eq:model}.
In particular, we employ Gaussian process regression (GPR)
\begin{equation}
\log\left( \kappab_0^{-1} \kappab(\rhob) \right)  = \tilde{\cs}^T \basis(\rhob) + \epsilon
\end{equation}
where $\basis = \{ 1, \rho_a, \rho_a \rho_b, \ldots \}$ is a vector of monomial basis functions modeling with coefficients $\tilde{\cs} = \{ \cb_a, \cb_{ab}, \ldots \}$, and $\epsilon$ is the noise in the mean due to statistical finite sampling errors.
GPR is an attractive representation since many of the necessary integral evaluations can be done analytically and yet it is a flexible model of the statistical variation.
Note we are using a polynomial basis to represent deviations from an assumed mean trend, in this case $\kappab \sim  \kappab_0 \exp(-\rhob)$, which is in-line with methods developed in \cref{kennedy2001bayesian}.

As described more fully in the Methods section, we collect data $\data \equiv [ \xdata, \ydata ] = [ \{ \log \kappab_{i} \}_I, \{ \rhob_{i}\}_I ]$ to calibrate $\tilde{\cs}$ at selected defect densities $\rhob_I$ over a number of replicates $\{\rhob_{i}\}_I$.
Given these observations, we obtain the (posterior) probability density of the coefficients $\posterior$ through Bayes' rule
\begin{equation}
\posterior = \frac{\likelihood \prior}{\evidence}
\end{equation}
$\likelihood$ is the likelihood of observing $\ydata$ given the coefficients $\coef$ and the densities $\xdata$,
$\prior = \Nc(0,\Sigmab)$ is the prior assumption of the distribution of the coefficients with covariance $\Sigmab$, and $\evidence$ is a normalizing factor.
The form of the likelihood 
\begin{equation}
\likelihood = \Nc( \xdata^T \coef, \tilde\Sigmab)
\end{equation}
is tied to assumptions about the error model.
In this case we assume a diagonal covariance $\tilde\Sigmab$ based on the observed variance in $\ydata$.
On the other hand, the covariance structure of the prior $\Sigmab$, which is related to the distribution of the coefficients, is given by a chosen covariance kernel $k$: 
\begin{equation}
\covar(\rhob_I, \rhob_J) \equiv \basis(\rhob_I)^T \As^{-1} \basis(\rhob_J)
= k\left(-\frac{\| \rhob_I - \rhob_J \|^2}{2 \ell^2} \right)
\end{equation}
(Here, we use a generalized radial basis function, the Mat\'ern kernel.)
The hyper-parameter $\ell$ is a correlation distance that is tuned to the data based on the (log) likelihood.
Finally, the posterior is
\begin{equation} \label{eq:posterior}
\posterior \propto \Nc( \As^{-1} \Bs, \As^{-1})
\end{equation}
with $\As = \xdata \tilde\Sigmab^{-1} \xdata^T + \Sigmab^{-1}$ and $\Bs = \xdata \tilde\Sigmab^{-1} \ydata$,
which leads to the model predictions:
$ \left( \As^{-1} \Bs \right)^T \bs(\rhob)$ for the mean and $ \basis^T(\rhob) \As^{-1} \basis(\rhob)$ for the associated covariance.
The mean is also the most likely / maximum \aposteriori (MAP) estimate in this model.
For further details consult  Rasmussen and Williams' text \cite{rasmussen2006gaussianCh2}.

\section{Method} \label{sec:method}

We employ an empirical potential to represent the interactions of B1 LiF crystals with a variety of defects.
The details of the atomistic model of LiF have been given in a previous publication \cite{jones2016estimates}.
Briefly, we use an empirical Tosi-Fumi/Born-Mayer-Huggins potential developed by Belonoshko \etal \cite{Belonoshko2000Born} for studying high pressure phase transitions.
It has the common long-range Coulomb plus short-range repulsive interactions form:
\begin{equation}
\Phi = \sum_{a\le b} \sum_{\alpha,\beta} \left( A_{ab} \exp\left(-B_{ab}r_{\alpha\beta}\right) + \frac{q_a q_b}{\epsilon_0 r_{\alpha\beta}^2} \right)
\end{equation}
where 
$r_{\alpha\beta}$ is the distance between atoms $\alpha$ and $\beta$, 
$a \in \{\text{Li}^+,\text{F}^-\}$ is the element type and $q_a$ is the charge of atom $\alpha$, 
and $\epsilon_0$ is the vacuum permittivity.
The Belonoshko \etal parameters are 
$\{ A_{\smallplus\smallplus}, A_{\smallplus\smallminus}, A_{\smallminus\smallminus} \} = \{  98.933, 401.319, 420.463 \}$ eV, 
$\{ B_{\smallplus\smallplus}, B_{\smallplus\smallminus}, B_{\smallminus\smallminus} \} = \{   3.3445,  3.6900,  3.3445 \}$ \AA$^{-1}$, 
with charges $\{ q_+, q_- \} = \{ +1, -1 \} e$.
We use a particle-particle particle-mesh (PPPM)\cite{hockney2010computer} long/short-range split method to solve the Poisson electrostatics problem for the Coulomb interactions.
For validation of the potential in the present context, defect energies calculated with the method in \cref{zhou2017atomistic} will be reported in the Results section.

By observing the equilibrium fluctuations in the system-wide heat flux $\Jb = \Jb(t)$ in these systems, 
the thermal conductivity tensor $\conductivityb$ can be obtained from the Green-Kubo formula:
\begin{equation}\label{eq:GK_conductivity}
\conductivityb \ = \ \dfrac{V}{k_B T^2} \int_0^{\infty} \Bigl< \heatfluxb(0) \otimes \heatfluxb(t) \Bigr> \, \dm t  \ ,
\end{equation}
where $V$ is the system volume, $T$ is the temperature, $k_B$ is the Boltzmann constant.
The bracket $\langle \cdot \rangle$ denotes the appropriate ensemble average, where $\langle \heatfluxv \rangle = \mathbf{0}$.
A formula \cite{Irving1950,Mandadapu2009} for the heat flux $\Jb$ suitable for MD is:
\begin{equation}
\heatfluxb = \frac{1}{V} \sum_\alpha \left( \varepsilon_\alpha \Ib + \boldsymbol{\nu}_\alpha^T \right) \vb_\alpha \ ,
\end{equation}
where the per-atom energy $\varepsilon_\alpha$ is formed from the kinetic energy of the atom $\alpha$ and a partition of the total potential energy $\Phi$  to individual atoms \cite{Schelling2002}, and the virial stress $\virial_\alpha$ for atom $\alpha$ in terms of the fundamental positions $\xb_\alpha$, velocities $\vb_\alpha$, and forces $\fb_\alpha$.
(The virial expression is described in detail in an appendix of \cref{jones2016estimates}.)

Interface conductance of planar defects can also be estimated directly via a GK method \cite{merabia2012thermal,chalopin2012thermal}
\begin{equation} \label{eq:G}
G  
= \frac{1}{k_B T^2 A} \int_0^\infty \left< \dot{H}_L(0) \dot{H}_L(t) \right> \mathrm{d}s 
\end{equation}
where $H_L(t)$ is the energy of atoms in region $L$ on one side of the planar defect (region $R$ being the other with $A$ being the area of the interface between them) and hence $\dot{H}_L$ is the total heat flux into region $L$ (from region $R$ since energy is assumed to be conserved).
Instead of direct evaluation of \eref{eq:G} using the expression in \cref{chalopin2012thermal}, since it involves forces on atoms in $L$ by all atoms in $R$ which is infeasible to evaluate with a long-range Coulomb interaction and a Poisson solver like PPPM, we use a central difference approximation:
\begin{equation}
\dot{H}_A (t_{n}) \approx \frac{1}{2 \Delta t} \left( H_A(t_{n+1}) - H_A(t_{n-1}) \right)
\end{equation}
for the change in energy of the region on one side of the planar defect, where $\Delta t$ is the time-step of the integration scheme and the subscript $n$ in $t_n$ denotes the index of the particular time.

Since we extract thermal conductivity from the molecular dynamics of defected LiF crystals using finite size, periodic cells, the range of representable densities and separations of the defects is limited.
To explore a wide range of defect densities we use 32$^3$ unit cells as a nominal system volume.
These large systems allows us to explore the coordination effects of neighboring defects on conductivity beyond that of periodic images.
Furthermore, the cubic B1 crystal structure of LiF restricts the types and relative orientations of defects such as dislocations.
As discussed in the Theory section, the probable distribution of defects can be characterized by densities and higher moments.  
We marginalize mean conductivities at specific densities over the higher moments by sampling over atomic arrangements relaxed to equilibrium.
Ten replica systems were used to estimate the mean thermal conductivity response and its variance for each defect density.
The details of how these configurations were obtained for each defect type examined will be given in the Results section. 
In general, perfect crystals were created at the compressed lattice parameter associated with the selected pressure (100 GPa), atoms were removed or rearranged to create the various defects at random locations, and then these systems where relaxed with isothermal, isobaric dynamics.

Attention was given to the possibility that thermalization and stress equilibration in isobaric dynamics might lead to the formation of additional defects or the coalescence of defects especially for the configurations with vacancies and dislocations.
The interactions of dislocations and vacancy point defects have already been rigorously explored \cite{bullough1970kinetics}. 
Here, we focus on the fact that a population of vacancies equal to the atoms in the extra half plane in a dislocation can result in annihilation of both.
The diffusion barriers for the vacancies and the related barriers for dislocation climb and glide, as well as any elastic costs, prevent this from happening instantaneously.
In preliminary studies we found that vacancy densities on the order of 7.5 nm$^{-3}$ and smaller in the presence of a dislocation density of 0.1 nm$^{-2}$ (which is effectively at the annihilation limit) were stable over the duration needed to run the GK simulations.
We did observe that in some cases the dislocation dipoles climbed which decreased the size of the extra half plane and indicated an accumulation of vacancies at the compressive side of the dislocations.

After construction and relaxation of $N_I=$10 initial structures with initial velocities sampled from the Boltzmann distribution for each defect density $\rhob_I$, the time correlation required by the GK expression, \eref{eq:GK_conductivity} was evaluated with constant energy dynamics with a 0.5 fs time-step and a 5 fs sampling interval for each replica.
We monitored relaxation of the heat flux $J(t)$ after the dynamics was switched from isobaric, isothermal to constant energy to obtain steady statistics for the GK correlation.
To estimate the ensemble average in \eref{eq:GK_conductivity} we average over constant temperature and constant defect density ensemble using random sampling to marginalize over uncontrolled degrees of freedom in phase space and higher spatial moments of the defect locations, as mentioned.
Then, using \eref{eq:model} as a model of the finite size, finite duration statistical errors in this mean, we apply Bayesian regression to calibration of the model:
\begin{enumerate}
\item Sample $\kappab(\rhob_I)$ using $N_I$ replicas averaged over $M_I$ steps at selected densities $\rhob_I$
\item Use the sample mean and variance of the replica data $\{\kappab_i\}_I$ at each point $\rhob_I$ to construct a likelihood with $\sigma_I = \sqrt{ \frac{ \var \{\kappab_i\}_I}{N_I}}$ being the expected error in mean $\bar{\kappab}_I$.
\item Then, with a prior based on perfect crystal data $\kappab(\mathbf{0})$, we sample the posterior, \eref{eq:posterior}, with a Markov chain Monte Carlo (MCMC) method and use kernel density estimation (KDE) to obtain a representation of the posterior distribution.
\end{enumerate}
The number of time samples $M_I$ used for each $\rhob_I$ vary with the computational cost of the particular system but at least 5 ns of simulation time was used for each.
Also the B1 cubic crystallographic structure renders $\kappab$ of undefected crystal isotropic and, hence, $N_I$=10 nominally isotropic replica systems with point defects yielded 30 samples of the components of $\kappab$.
Systems with (parallel) dislocations or planar defects break this isotropy leading to 20/10 and 10/20 samples, respectively, of the components in the perpendicular and parallel directions, respectively.

With this model we predict the thermal conductivity $\kappab$ of a large, crushed system and compare a direct, independent GK estimate of $\kappab$.
To extract density of defects in the crushed crystal (and to check the post-relaxation defect densities of the systems with pre-defined defects) we use the DXA algorithm \cite{stukowski2009visualization}.
The DXA unambiguously identifies dislocation lines using a bond-topology scheme, and, as by-product, can identify atoms in a non-crystalline environment.
By clustering these we identify point vacancies using a pre-conceived range of how many atoms should surround a vacancy.
(This scheme can fail to identify merged or closely neighboring vacancies.)

\section{Results} \label{sec:results}

After a preliminary validation of the defect energies predicted by the chosen potential, we calculated the thermal conductivity of systems with vacancies, dislocations, and their combinations in order to calibrate the thermal conductivity model $\kappab = \kappab(\rhob)$, \eref{eq:model}.
Then we compare the prediction versus a direct estimation of thermal conductivity of a large crushed system via the same GK methods.
Finally, we explore the effect of planar tilt boundaries on the thermal conductivity of LiF.

For this study, we chose 100 GPa, 1000 K as representative conditions for the ramp compression experiments with an average compressed lattice constant $a$ = 3.483 \AA\, for a perfect crystal with B1 unit cell.
(These conditions are approximately the center of the study reported in \cref{jones2016estimates}.)
At these conditions, the conductivity of a perfect crystal changes slightly with system size:
$\kappa=$
 27.0$\pm$0.5 W/m-K at $(4a)^3$,
 30.6$\pm$0.6 W/m-K at $(8a)^3$,
 30.2$\pm$0.5 W/m-K at $(16a)^3$, 
 30.2$\pm$0.4 W/m-K at $(32a)^3$, 
and effectively converges for $V\ge (8a)^3$.

\subsection{Defect energies}
In previous work \cite{jones2016estimates}, we compared the phonon properties of the chosen potential with \abinitio density functional theory (DFT) estimates; here, we calculate the defect energies as a means of validating it for this study.
We considered two types of vacancies, a collocated (covacancy) and a separated Schottky Li-F pair vacancy, as well as a $a/2$\miller{110} edge dislocation dipole.
The energies are determined with long time (32 ns) averages at low temperature ($\approx$1K) based on the work of Zhou \etal \cite{zhou2017atomistic} which showed that this approach leads to estimates that are superior to energy minimization techniques.

\begin{figure}[h!]
\centering
\subfigure[]
{\includegraphics[width=\figwidth]{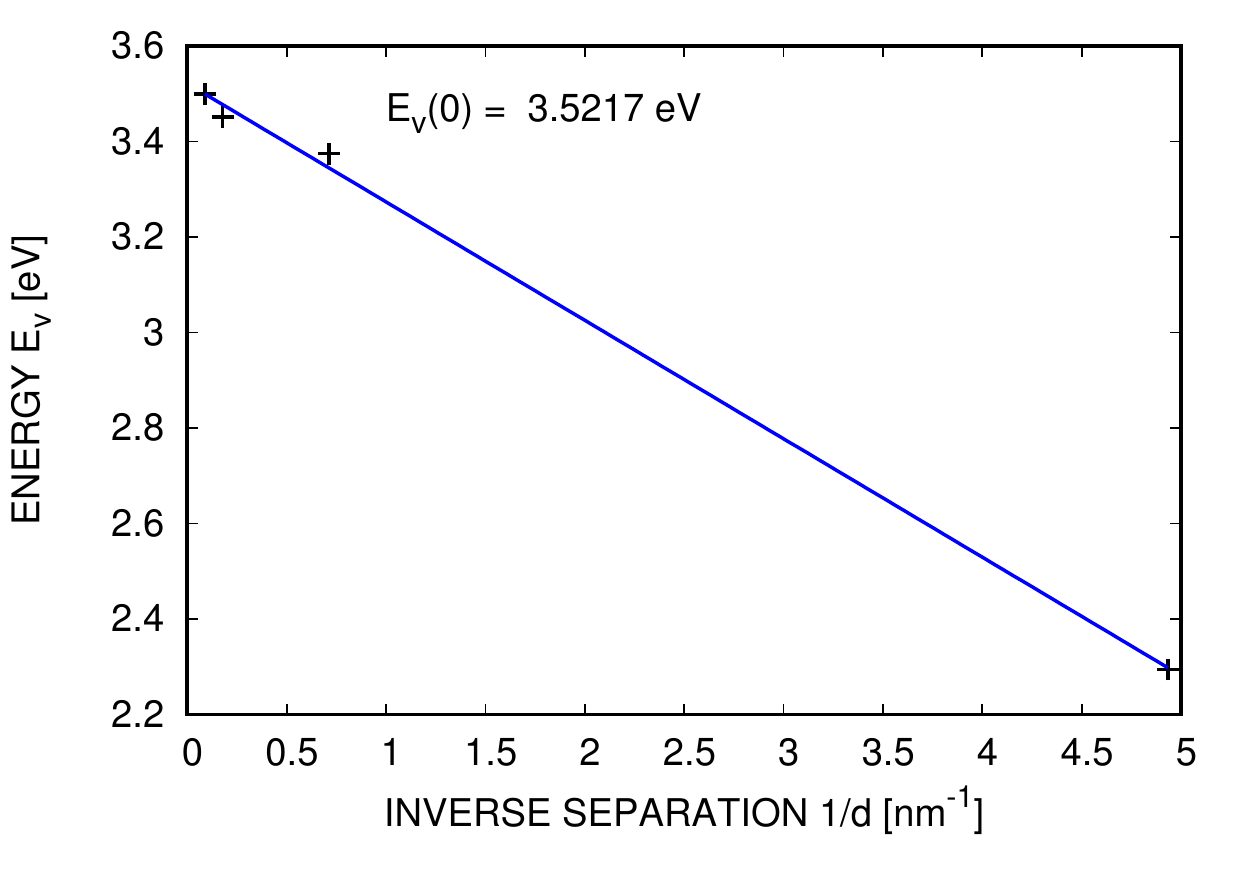}}
\subfigure[]
{\includegraphics[width=\figwidth]{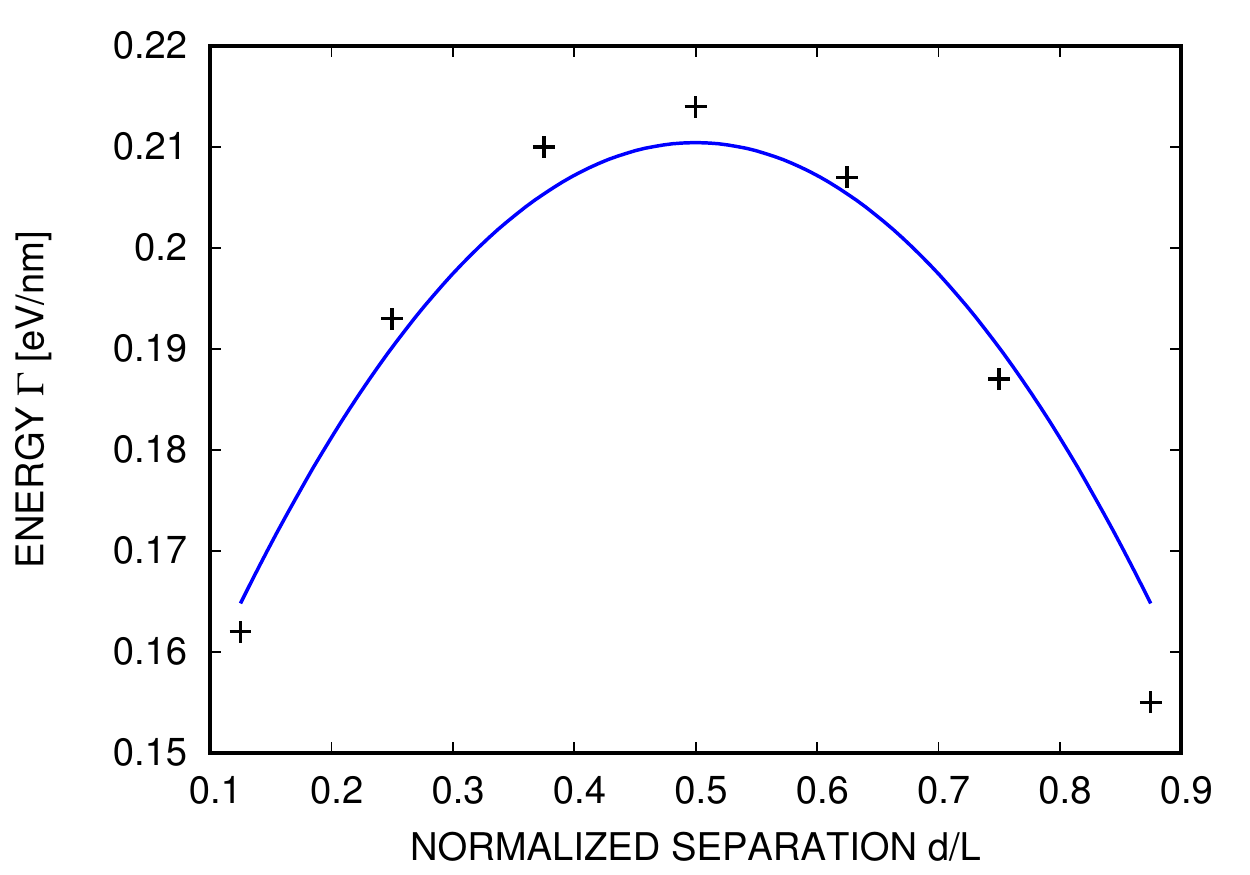}}
\caption{Defect energy: (a) Schottky divacancy as a function of the inverse separation distance, (b) line energy for a $a/2$\miller{110} dislocation dipole as a function of the dislocation core separation relative to the simulation box length aligned with the dipole (the core energy and an elastic dependent on the magnitude of the Burgers' vector determine the maximum, see \cref{zhou2017atomistic} for details).
}
\label{fig:defect_energies}
\end{figure}

Since the cohesive energies we obtained were slightly size dependent (likely due to the PPPM solver), we calculated the vacancy energies for system sizes $(4a)^3$, $(8a)^3$, $(16a)^3$, and  $(32a)^3$ relative to the cohesive energy obtained for the same size system.
Using this procedure, we obtained consistent defect energies of 2.28 eV at zero pressure and 3.43 eV at 100 GPa for the covacancy where a neighboring Li-F pair (separated by $a$/2) was removed.
The energy of Schottky vacancies separated by $\approx \sqrt{3} L/2$, where $L$ is the system box length, show a slight dependence on system size, see \fref{fig:defect_energies}a.
Motivated by the Coulomb energy of a point charge dipole, we fit this data to  $E_v(d) = C_0 + C_1/r$ and extracted 3.52 eV for the divacancy energy at infinite separation. 

We also performed energy minimization DFT simulations under the local density approximation of a small (4a)$^3$ cell at zero pressure and 100 GPa with 2x2x2 k-points, and a 800 eV energy cutoff.
We obtained we obtained cohesive energies, -5.25 and -1.67 eV/atom, respectively, at the two pressures using a 512 atom perfect lattice cell; and, then, calculated defect energies, 1.50 and 6.66 eV/divacancy, respectively, using the difference in energy of  510 atom cells and the corresponding cohesive energies.
Despite the disagreement between MD and DFT values and with experimental results\cite{haven1950ionic,stoebe1967ionic,barsis1967ionic} (2.68, 2.34, 2.40 eV at ambient conditions) the increased enthalpic costs with increased pressure is notable.
The primary influence of these discrepancies is in the formation of defects in the rapid compression simulation.
Also relevant to the fidelity of the fixed-charge, empirical potential we employed, we calculated the charges associated with the nuclei in the defected systems using Bader analysis and determined that they do not deviate more that 5\% from $\pm$1$e$.

In addition to the vacancy energies, we calculated the dislocation energy for a $\left< 110 \right>$ dislocations dipole in a periodic cell using a technique outlined in \cref{zhou2017atomistic}.
At ambient conditions these dislocations have a Burgers' vector $b$= 2.865 \AA.
By varying the separation of the parallel dislocations in the dipole the energies can be compared to linear elastic solution, see \fref{fig:defect_energies}b, and the core energy can be determined separately from the perturbation elastic energy.
Since the dislocation cores are effectively charge neutral no correction was made to the elastic solution.
(Details on how these systems were created will be given in the Results section.)
The fit to the analytical series solution in terms of the ratio of the distance between the dislocation cores to the periodic box dimension in that direction yielded a core energy of 1.0 eV/\AA, assuming a core radius of 2$b$.

Despite that our primary interest is in a non-equilibrium process that generates defects, we also calculate equilibrium defect concentrations as a point of reference and to provide an initial condition for the rapid compression simulation.
Well-established theory for equilibrium concentration point defects \cite{franklin1972statistical,KittelCh18} gives the mole fractions of the atomic vacancies comprising a Schottky pair defect $ x_+ = x_- = \exp\left(-\frac{G_S}{2 k_B T}\right) $
in terms of its formation energy $G_S$, Boltzmann's constant $k_B$ and temperature $T$.
Using experimental values for $G_S$, we obtain
$x_\pm \approx 10^{-21}$ at ambient temperature and pressure, 
$x_\pm \approx 10^{-6}$ at 1000 K and ambient pressure, and 
$x_\pm \approx 10^{-11}$ at 1000K, 100 GPa accounting for pressure-volume work.

No corresponding theory exists to predict the density of dislocations at ambient conditions since the expectation at equilibrium is that dislocations annihilate or move to the surface  but experimental measurements exists.
In LiF the observed dislocation density ranges from 
10$^{9}$ to 10$^{10}$ m$^{-2}$ at ambient conditions \cite{johnston1959dislocation,thacher1967effect,gupta1975dislocation,gilman1960behavior} and up to 10$^{14}$ m$^{-2}$ for deformed crystals\cite{andreev1968formation}.

\subsection{Point defects} \label{sec:vacancies}
To explore the effect of point defects on thermal conductivity, we constructed systems with populations of Schottky divacancies by removing an equal number of Li and F from periodic (32$a$)$^3$ perfect crystals at the compressed 100 GPa lattice spacing, energy minimizing and then equilibrating at the selected pressure and temperature using isobaric, isothermal dynamics.
We simulated four point defect densities with  $\{2^1,2^4,2^7,2^{10}\}$ vacancies using 10 replica systems each.
The resulting lattice constant was identical to the perfect crystal at the lowest selected vacancy density, 0.00144 nm$^{-3}$ (2:8$\times$32$^3$), and 0.1\% smaller at highest vacancy density, 0.738 nm$^{-3}$  (2$^{10}$:8$\times$32$^3$).
Also, by examining  a histogram of cluster sizes of atoms without B1 coordination, we verified that the initial vacancy density was preserved in the systems used to calculate the heat flux correlations.
We found that the resulting vacancy density was insensitive using a range of clusters sizes centered on 12; in particular, we counted non-crystalline clusters with sizes 12$\pm 4$ as point defects.

\begin{figure}[h!]
\centering
\subfigure[]
{\includegraphics[width=\figwidth]{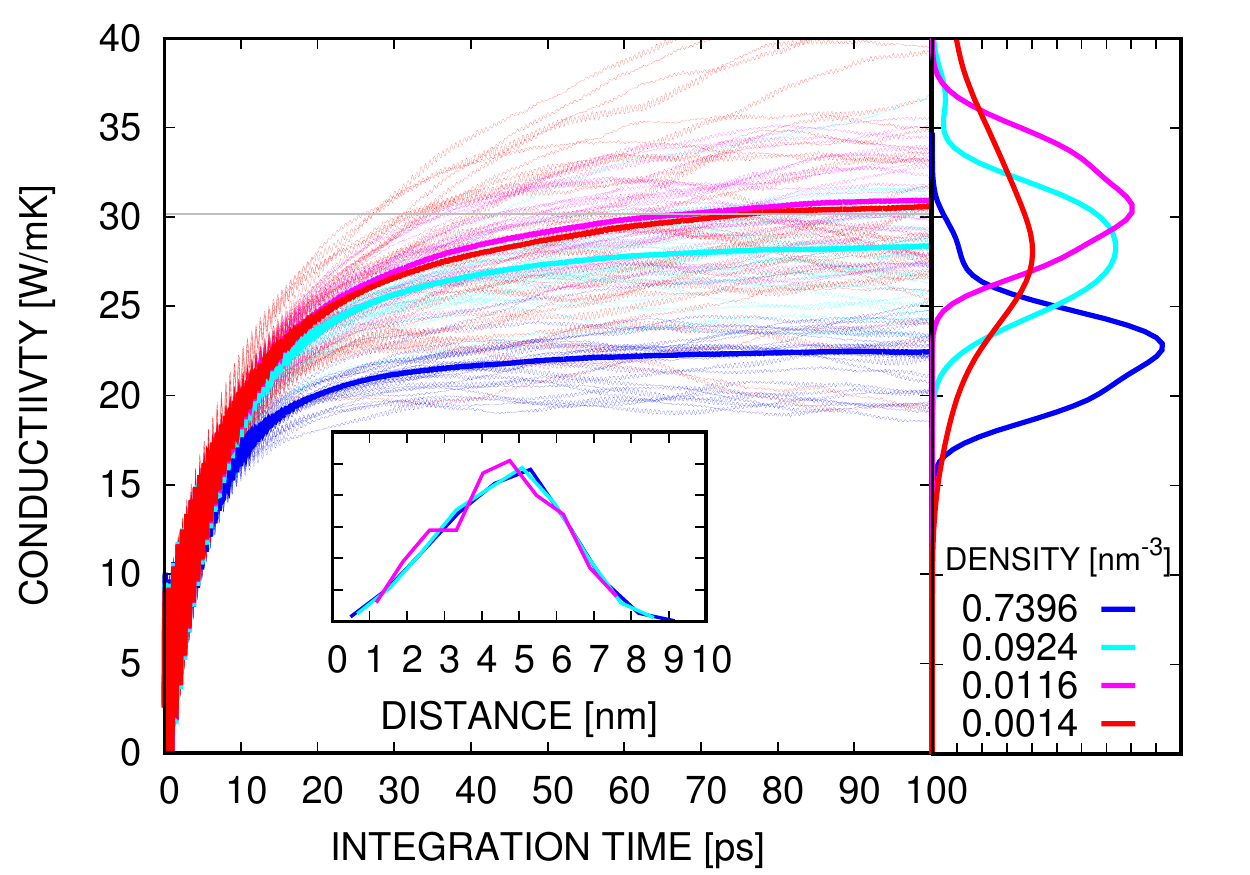}}
\subfigure[]
{\includegraphics[width=\figwidth]{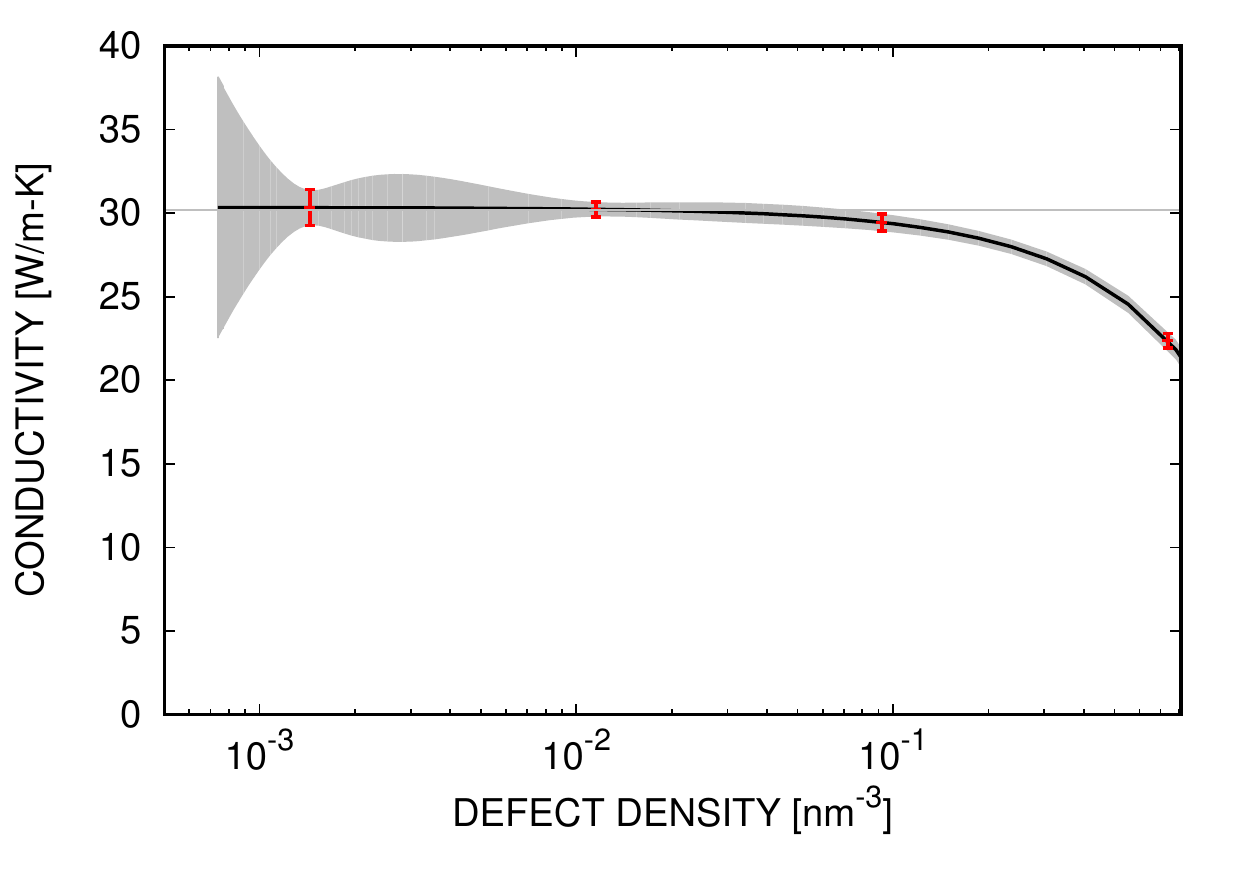}}
\caption{Thermal conductivity of LiF with divacancies: (a) correlation integrals (color denotes divacancy population density, thick color lines are the averages for each density, and the gray trend-line is the expected value for a perfect crystal), and inset: distance distribution (for concentration higher than 0.002 nm$^{-3}$), and
(b) thermal conductivity as a function of divacancy concentration (red: sample mean and one standard deviation error of the data, black: mean prediction, and gray: one standard deviation prediction of the model).
}
\label{fig:vacancy_conductivity}
\end{figure}

The inset of \fref{fig:vacancy_conductivity}a shows the distribution of pair-wise distances of vacancies identified in all the replicas with the cluster analysis post equilibration.
As can be seen, the distribution is effectively constant for the higher defect densities implying that the spatial distribution of vacancies is uniform.
\fref{fig:vacancy_conductivity}a shows the distribution of flux correlation integrals as a function of the upper limit of the integral in \eref{eq:GK_conductivity} for each of the 30 replicas at each of the selected vacancy densities.
The distributions resemble normal distributions and their widths are correlated with the amount of averaging time allowed for calculation of the flux correlations (the lowest density was run for less time, and hence broader, than the highest density).
Despite the systematic trend in the average values, refer to \tref{tab:kappa_point_line} and the thicker lines in \fref{fig:vacancy_conductivity}a, there is clearly overlap in the distributions of the 30 samples per $\rhob_I$. 
The GPR model of this data, shown in \fref{fig:vacancy_conductivity}b, indicates that for vacancy densities less than $10^{-2}$ nm$^{-3}$ the conductivity is effectively that of a perfect crystal to within error.
It is also noteworthy that the GPR model clearly shows the rapid increase in predicted error when using the model for extrapolation.
A similar prediction was achieved in \cref{zhou2009towards} using Monte Carlo sampling.

\subsection{Line defects}
\begin{figure}[h!]
\centering
{\includegraphics[width=1.2\figwidth]{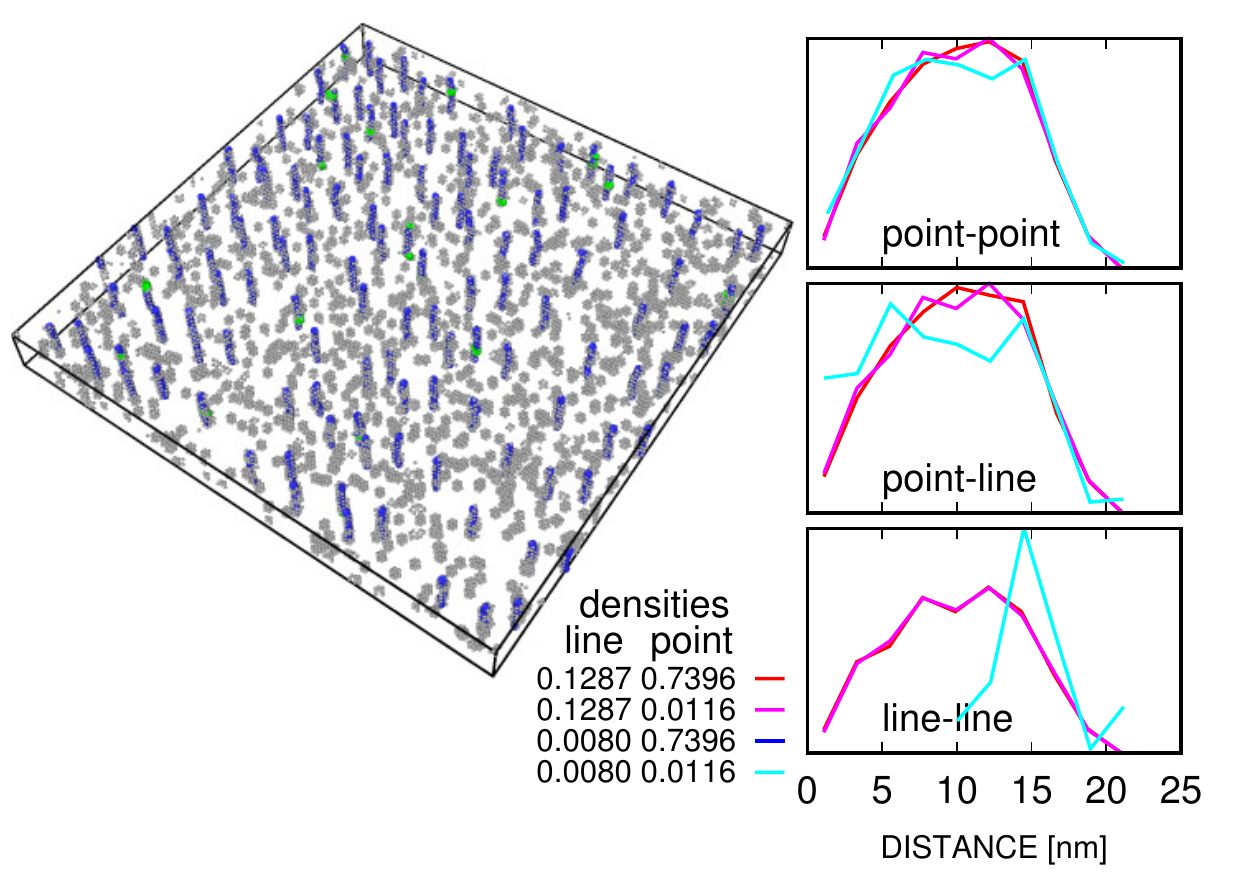}}
\caption{Vacancy-dislocation system at highest vacancy and dislocation density showing vacancies as clusters of not fully coordinated (gray) atoms and dislocations as (blue, parallel) lines, and histograms of pair-wise defect distances.
} 
\label{fig:vacancy-dislocation}
\end{figure}

To explore the effect of line defects on thermal conductivity, we constructed replica systems with $\{2^1,2^3,2^5,2^7\}$ dislocations by removing neighboring partial \millersurface{$1\bar{1}0$} Li,F planes of random extent less than half box dimension from a (64$a\sqrt{2}$)$^2\times$(8$a$) compressed perfect lattice, energy minimizing this structure, then thermalizing it with isobaric, isothermal dynamics.
The resulting systems were aligned with the \miller{110}, \miller{110}, \miller{001} crystal directions and all the parallel dislocations have $a/2$ \miller{110} Burgers vector and \miller{001} line direction. 
For simplicity and computational efficiency, we kept the cell dimension in dislocation line direction relatively small (see  \fref{fig:vacancy-dislocation} for a representative configuration with additional point defects).
This arrangement did not encourage dislocation entanglement, which will be discussed in more detail in the next section.
The resulting lattice constant was effectively unchanged at the lowest dislocation density 0.0020 nm$^{-2}$ and 0.8\% smaller in-plane for the highest line density 0.1287 nm$^{-2}$.

\begin{figure}[h!]
\centering
\subfigure[]
{\includegraphics[width=\figwidth]{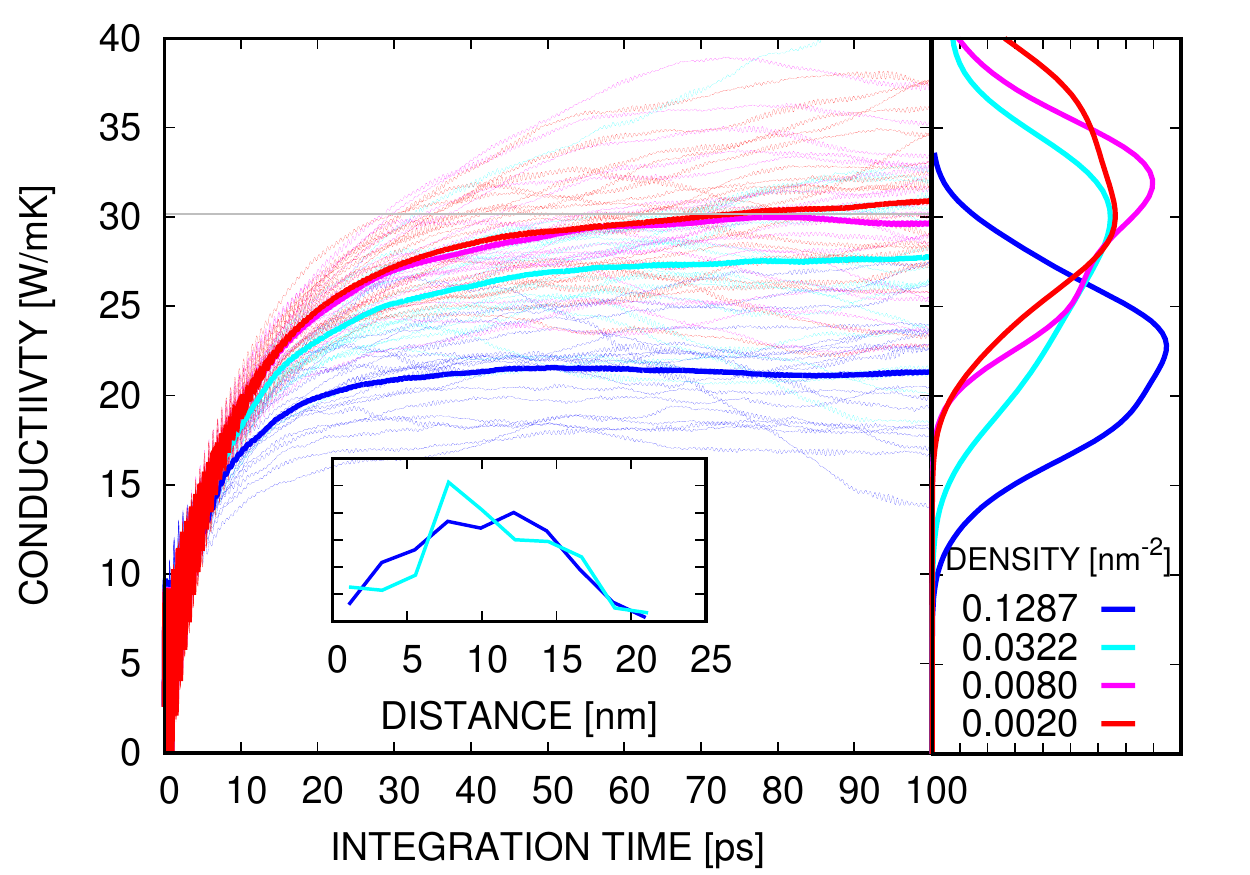}}
\subfigure[]
{\includegraphics[width=\figwidth]{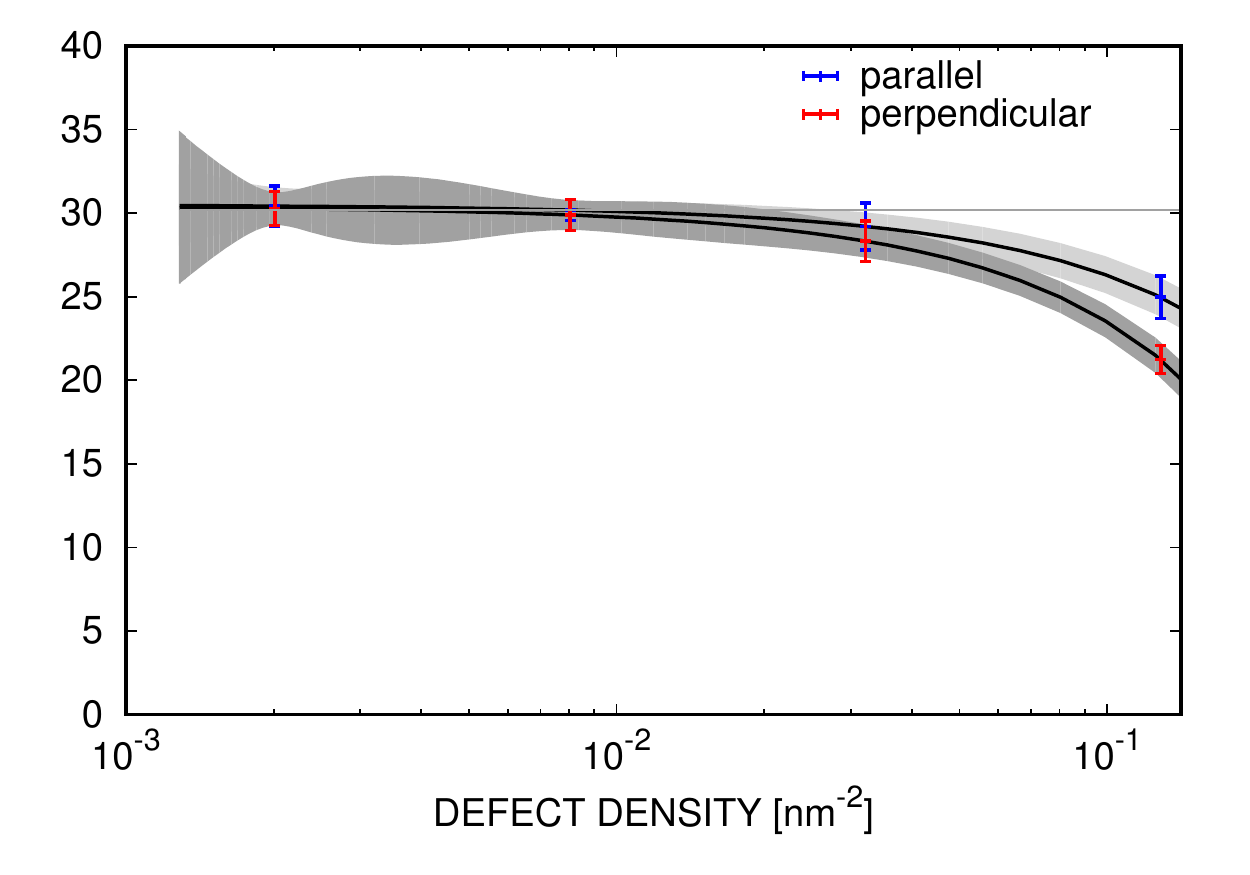}}
\caption{Thermal conductivity of LiF with \miller{110} dislocations: (a) correlation integrals for conduction perpendicular to the dislocation lines (color denotes divacancy population density, thick color lines: average for a specific line density, and  gray trend-line:perfect crystal), and inset: distance distribution (for line density higher than 0.002 nm$^{-2}$), and
(b) thermal conductivity as a function of dislocation concentration for conduction perpendicular and parallel to the lines (red and blue: sample mean and one standard deviation error of the data, black: mean prediction, gray: one standard deviation prediction of the model).
}
\label{fig:dislocation_conductivity}
\end{figure}

The inset of \fref{fig:dislocation_conductivity}a shows the distribution of pair-wise distances of between the parallel dislocations across all the replicas post equilibration.
The distributions appear to be converging, albeit more slowly than in the point defect case most likely due to the fewer numbers of dislocation lines.
\fref{fig:dislocation_conductivity}a shows the flux correlation integrals for 20 samples (10 replicas $\times$ 2 equivalent directions perpendicular to the dislocation lines).
Clearly dislocation densities higher than 0.03 nm$^{-2}$ are necessary to have a significant effect on conductivity.
This is born out in the GPR models for both the perpendicular and parallel conduction, \fref{fig:dislocation_conductivity}a.
The effect on conductivity parallel to the lines is also significant, albeit of a lesser magnitude than the effect perpendicular to the lines.
Note that, due to the geometry of the simulation cells, only 10 samples of the parallel conduction data was available at each density and this is reflected in the larger error band.

\begin{figure}[h!]
\centering
\subfigure[]
{\includegraphics[width=\figwidth]{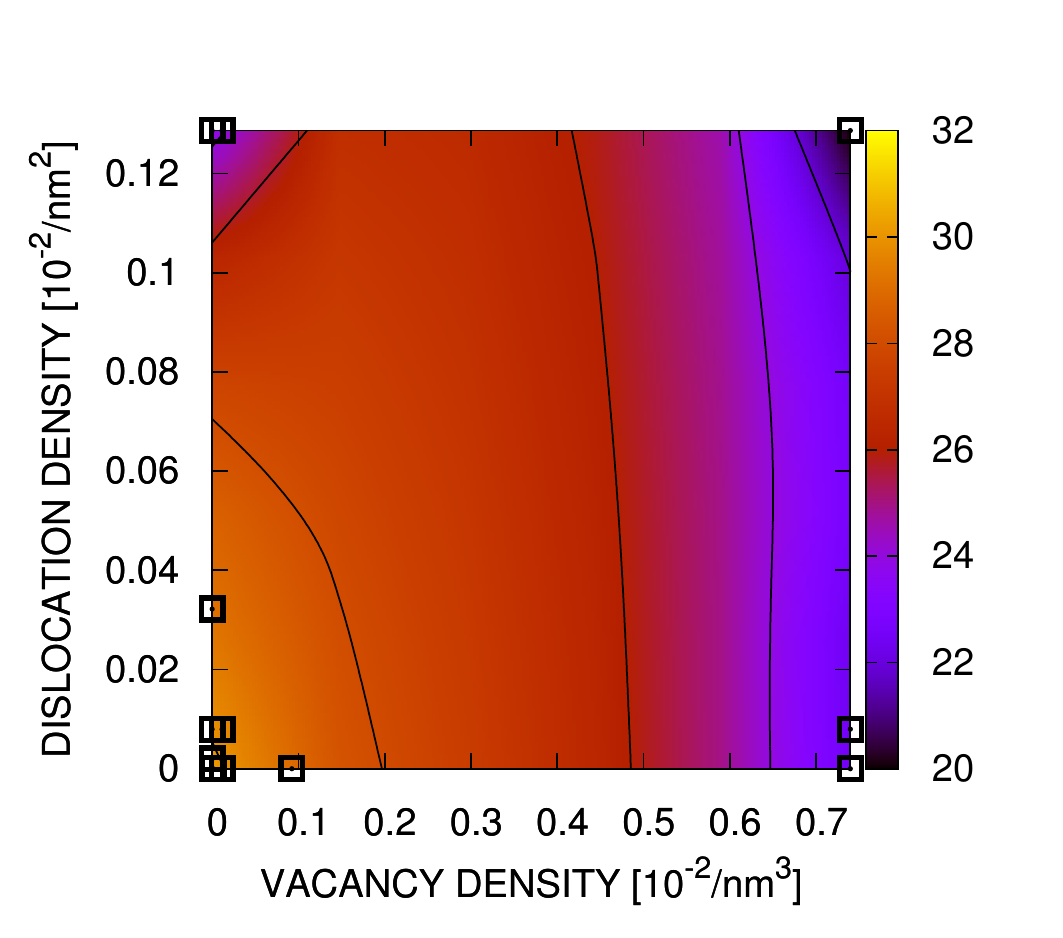}}
\subfigure[]
{\includegraphics[width=\figwidth]{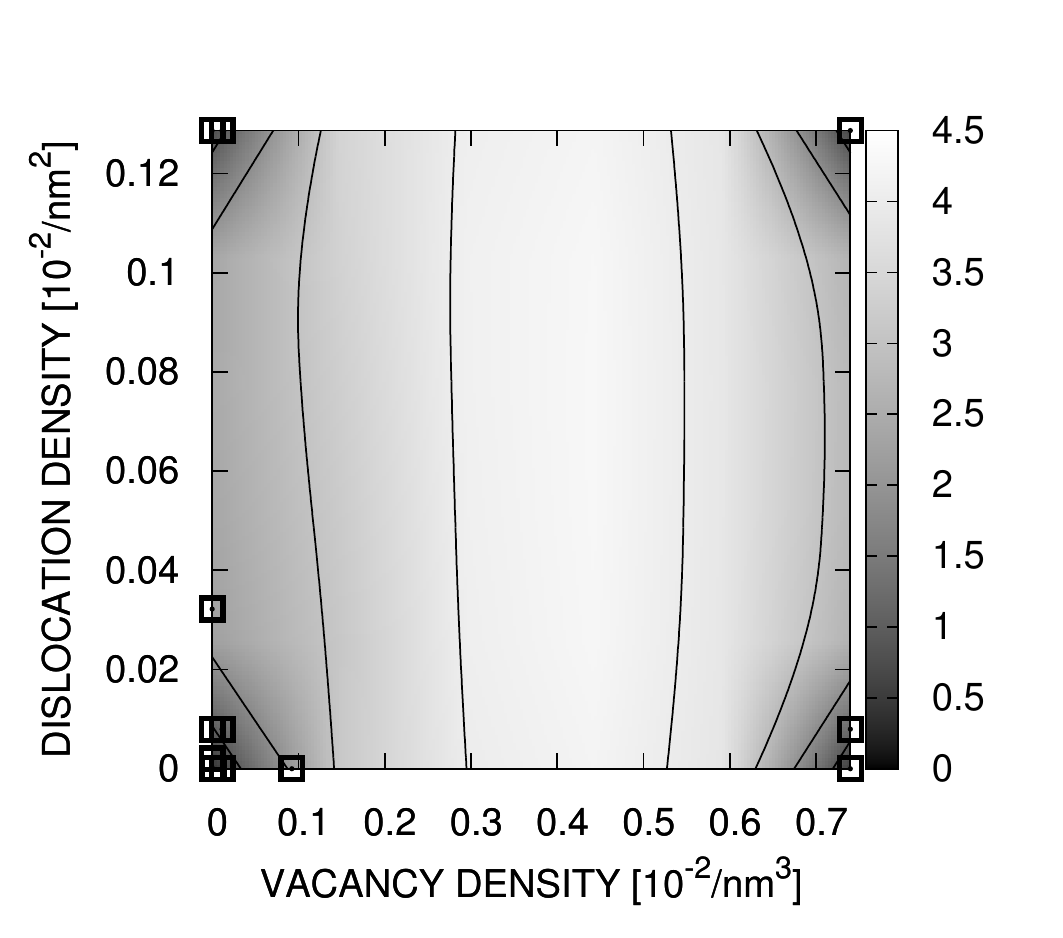}}
\caption{Vacancy-dislocation: (a) conductivity perpendicular to the dislocation lines and (b) error in conductivity. 
Points indicate GK sample locations.
}
\label{fig:vacancy-dislocation_conductivity}
\end{figure}

To quantify the joint of effects of vacancies and dislocations we created vacancies in selected systems with dislocations, re-equilibrated them and calculated thermal conductivities using the same procedures.
A sample configuration of the highest dislocation and vacancy density is shown in \fref{fig:vacancy-dislocation} and the insets show that the spatial correlations of the defects are behaving as in the previous studies with just vacancies or dislocations.
Table \ref{tab:kappa_point_line} summarizes the conductivity results for the full set of vacancy and dislocation densities we explored.
Clearly, the data shows a monotonic decrease in conductivity with increasing density of both vacancies and dislocations.
Since we mean to apply this data to a simulation with no constraint on the dislocations being aligned, we created a GPR model of the combined effects using a simple homogenization of the line-parallel $\kappa_\|$ and line-perpendicular $\kappa_\perp$ conductivity:
\begin{equation} \label{eq:ave_kappa}
\bar{\kappab}
= \int_{S^2} \kappab_\nb \probability(\nb) \, \mathrm{d}S
= \frac{1}{3} \tr \kappab 
\end{equation}
assuming $\kappab$ is diagonal ($\kappab_\nb = \kappa_\| \nb \otimes \nb + \kappa_\perp (\Ib - \nb \otimes \nb)$) and $\probability(\nb)$ on the unit sphere $S^2$ is uniform so that $\int \nb \otimes \nb \, \mathrm{d}S = 1/3\, \Ib$.
Using $1/3 ( \kappa_\| + 2 \kappa_\perp )$ as the average conductivity of LiF with a randomly oriented network of dislocations (and vacancies), we constructed the GPR model of their joint effect on thermal conductivity shown in \fref{fig:vacancy-dislocation_conductivity}.
Overall the model displays the expected trend with increased defect density; however, the sparse, log sampling of densities leads to regions of high uncertainty $>$ 10\%.

\begin{table}
\centering
%%%%%%%%%%%%%%%%%%%%%%%%%%%%%%%%%%%%%%%%%%%%%%%%%%%%%%%%%%%%%%%%%%%%%%%%
\begin{tabular} {|c | c | c | c | c | c |}
\hline
vacancy & \multicolumn{5}{|c|}{dislocation {[}nm$^{-2}${]}}\\
 {[}nm$^{-3}${]}  & 0  & 0.0020  & 0.0080  & 0.0322  & 0.1287 \\
\hline
\hline
  0  &  30.18 $\pm$ 0.37  &  30.89 $\pm$ 1.02  &  29.63 $\pm$ 0.92  &  27.79 $\pm$ 1.21  &  21.30 $\pm$ 0.83 \\
     &                    &  30.65 $\pm$ 1.20  &  29.96 $\pm$ 0.63  &  30.20 $\pm$ 1.41  &  24.84 $\pm$ 1.27 \\
\hline
 0.0014  &  30.59 $\pm$ 1.07 &  &  &  &  \\
            &         &  &  &  &  \\
\hline
 0.0116  &  30.92 $\pm$ 0.45 &  &    
25.84 $\pm$ 2.25 &  &    20.81 $\pm$ 0.82 \\
         &                   &  &    
31.05 $\pm$ 2.17 &  &    22.17 $\pm$ 0.65 \\
\hline
 0.0924  &  28.35 $\pm$ 0.52 &  &  &  &  \\
         &         &  &  &  &  \\
\hline
 0.7396  &  22.44 $\pm$ 0.44 &  &    
21.30 $\pm$ 0.82 &  &    19.96 $\pm$ 0.35 \\
         &                   &  &    
20.36 $\pm$ 0.43 &  &    20.11 $\pm$ 0.59 \\
\hline
\end{tabular}
%%%%%%%%%%%%%%%%%%%%%%%%%%%%%%%%%%%%%%%%%%%%%%%%%%%%%%%%%%%%%%%%%%%%%%%%
\caption{Thermal conductivity (in W/m-K) as a function of point (divacancies) and line ($\langle 100 \rangle$ dislocation) defect densities.
For systems with dislocations, the upper entry in the table is the conductivity perpendicular to the dislocation lines and the lower entry is the conductivity parallel to the lines.
}\label{tab:kappa_point_line}
\end{table}


\subsection{Rapidly compressed crystal}

To create a LiF sample representative of the high-strain rate experiments we create a $(64a)^3$ single crystal with a single divacancy (2,097,150 atoms) and thermalized it at ambient pressure and $T$=1000 K.
The single pair vacancy gives a pre-existing defect concentration on par with the equilibrium density at the selected temperature (but far in excess of that at ambient temperature) and was also lowest concentration possible.
We then rapidly compressed this system in the \miller{100} direction to to 100 GPa in 10 ps (strain rate $\approx$ 10$^{9}$ s$^{-1}$) to mimic the experimental ramp compression.
We applied zero strain in lateral directions to mimic inertia confinement.
The resulting compressive stress state was 100 $\eb_1 \otimes \eb_1 +$ 97 $(\Ib - \eb_1 \times \eb_1)$ GPa at 0.346 $\eb_1 \otimes \eb_1$ strain.
Admittedly this method of creating the system deviates from the actual experimental conditions but was necessary to create a periodic cell suitable to apply the GK method.

\begin{figure}[h!]
\centering
\subfigure[]
{\includegraphics[width=0.7\figwidth]{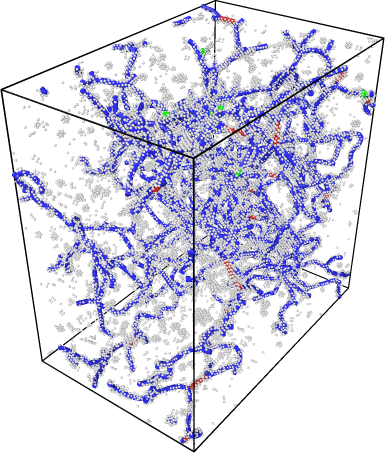}}
\subfigure[]
{\includegraphics[width=\figwidth]{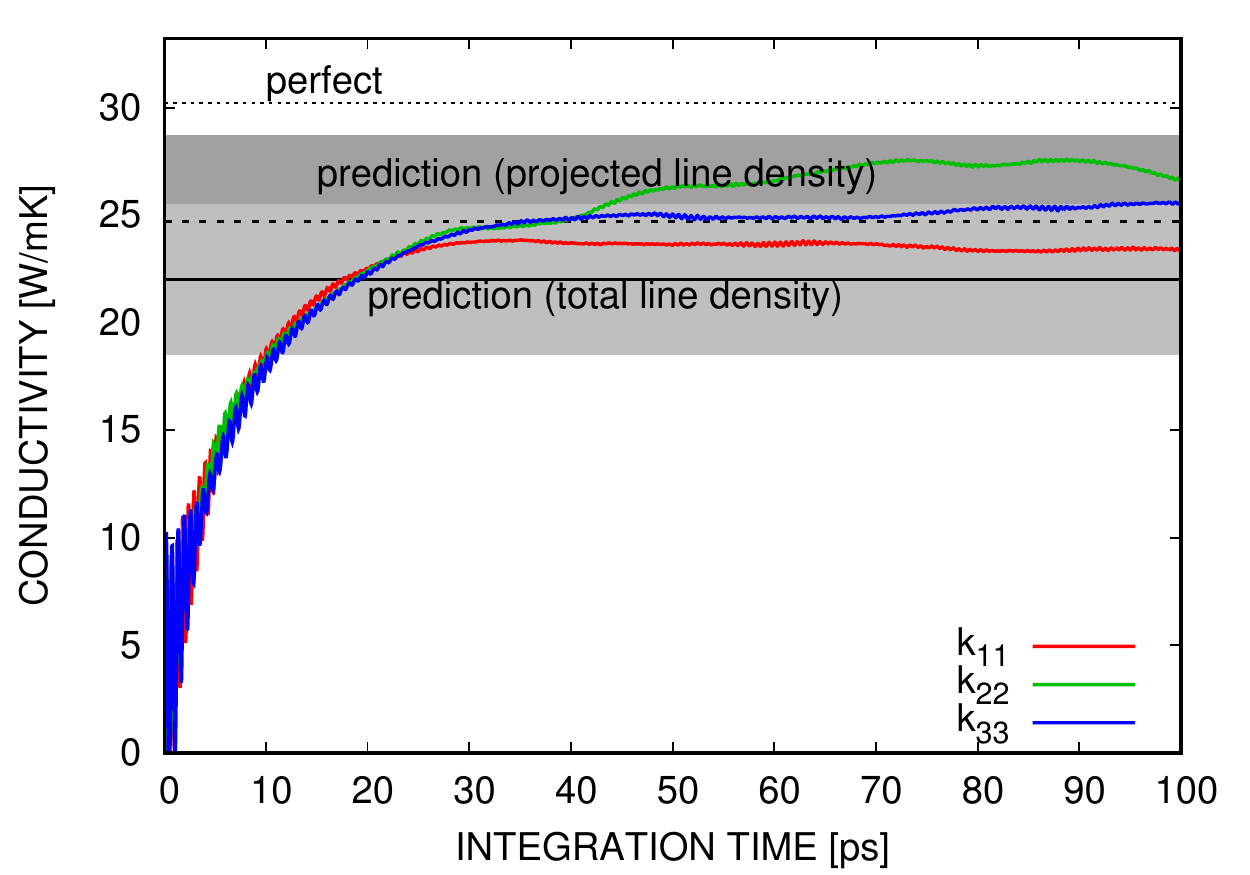}}
\caption{Rapid compression: (a) dislocations (blue, density: 0.1231 nm$^{-2}$) and defect clusters (gray, density: 0.01229 nm$^{-3}$) at a diagonal stress state (100, 97, 97) GPa with corresponding strains: (34.5, 0.0, 0.0) \%
(b) thermal conductivity (color line: correlation integrals for the components of $\kappab$ from direct GK calculation, dotted line: $\kappab$ for a perfect crystal at 100 GPa pressure, solid line and light gray error band: model prediction using estimated point vacancy density and total line length density, dashed line and dark gray error band: model prediction using estimated point vacancy density and projected line length density)
}
\label{fig:crush}
\end{figure}

Since in this case the defects were not pre-determined, we relied heavily on DXA algorithm \cite{stukowski2009visualization,stukowski2010extracting} to identify dislocations and the derivate tuned cluster method described in \sref{sec:vacancies} to identify vacancies defects.
This analysis of the final configuration shown in \fref{fig:crush}a gave a point defect density of 0.01229 nm$^{-3}$ and a line defect density of 0.1231 nm$^{-2}$, 
(Since we have no quantitative way of characterizing surface defect or measuring their area and hence their density, we consider them separately in the next section.)

A direct application of the GK method to this system gave the correlation integrals shown in \fref{fig:crush}b which are apparently converged and display anisotropy $\kappa_{11} < \kappa_{33} < \kappa_{22}$.
Prediction of the GPR model, \fref{fig:vacancy-dislocation_conductivity}, is also shown in \fref{fig:crush}b as the solid black with gray error band.
The prediction encompasses the direct $\kappa_{11}$ value albeit with marginal correspondence.
If we instead use the projected line density in the compressed direction, 0.0628 nm$^{-2}$, instead of the total line density, we bring the prediction (dashed line) closer to the direct measurement.
(The projected line density is 
$\rho_i = \int \left\| \mathsf{P}_i \frac{\mathrm{d}\xb(s)}{\mathrm{d}s} \right\| \, \mathrm{d}s $,
where $\mathsf{P}_i$ is the projection operator for plane with normal $\eb_i$.)
Also, the projected line densities, $\rho_{2,3}$ = 0.0588, 0.0610 nm$^{-2}$, correlate with the ordering of the  directional components $\kappa_{22,33}$ of the conductivity tensor which are nominally at the same compression. 
We believe this is an indication that conductivity is sensitive to dislocation structure in ways beyond those accounted for in the model we calibrated.

\subsection{Planar defects}

We examined both the conductivity of LiF with \millersurface{210} tilt boundaries and the intrinsic conductance of these interfaces.
The four ($n a\sqrt{5}$)$\times$(8$a\sqrt{5}$)$\times$(8$a$) systems, with $n$=12, 24, 48, 96, were  oriented \miller{210} \miller{$\bar{1}$20} \miller{001} and had two tilt boundaries to allow for periodicity.
In this study, unlike the previous point and line defect explorations,  we scale the size of the system for efficiency which may have incurred simulation size effects in addition to the physical effects of defect density; however, even the smallest system is larger than that needed to avoid size effects in a perfect crystal.

The thermal conductivity perpendicular and parallel to the tilt boundary was estimated with the same GK formula, \eref{eq:GK_conductivity}, used in the point and line defect cases.
These results are summarized in \tref{tab:kappa_interface}. 
They show a slow convergence to the perfect lattice conductivity with decreasing tilt density and a distinct effect of the tilt boundary on conduction parallel to the interfaces.
Note that the simple homogenization technique, \eref{eq:ave_kappa}, also applies in this case.

\begin{figure}[h!]
\centering
{\includegraphics[width=\figwidth]{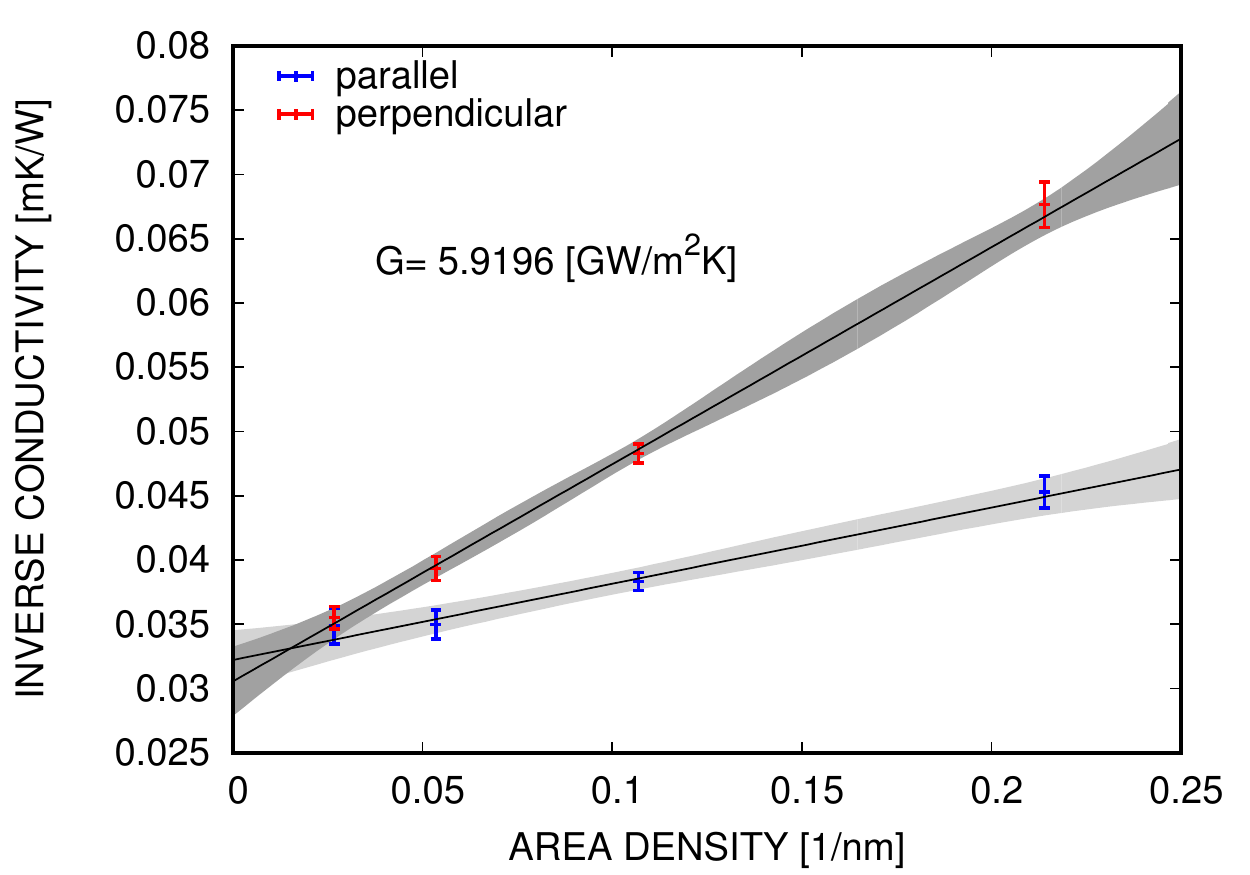}}
\caption{Tilt boundary conductivity plotted as $1/\kappa$ as a function of density for both the conductivity perpendicular and parallel to the \millersurface{210} defect surface. 
The interface conductance $G$ is recovered from the perpendicular conductivity data.
}
\label{fig:tilt_conductivity}
\end{figure}

From flux conservation we can arrive at a version of Matthiessen's rule for independent scatterers
\begin{equation}\label{eq:GL}
 \frac{1}{\tilde{\kappa}} = \frac{1}{\kappa_\infty} + \frac{\rho}{G}
\end{equation}
that relates $\tilde{\kappa}$, the apparent/effective conductivity of the whole cell with the interface, to  $\kappa$ the corresponding conductivity without the interface resistance but with the distortions of the interface on the lattice, and $G$, the conductance of the interface.
Note with parallel interfaces $\rho$ is equal to the inverse of the normal interface spacing.
Using regression we can extract both the conductivity normal to the interfaces $\kappa_\infty \approx 30.3$ W/mK and $G = 5.92$ GW/m$^2$K from a sequence of $\tilde{\kappa}$ values at different densities $\rho$, as \fref{fig:tilt_conductivity} shows.
This limiting conductivity is consistent with the perfect crystal value as expected.
The value of the interface conductance is comparable to those (1--10 GW/m$^2$K) reported in \cref{chalopin2012thermal} and \cref{landry2009effect} for a Si/Ge super-lattice.
As mentioned in the Theory section, we also used a GK method, \eref{eq:G}, to directly calculate $G$.
For $\rho$= 0.0535 nm$^{-1}$ (normal spacing 18.7 nm) systems, we obtained a significantly lower value $G$ = 0.28 $\pm$ 0.20 GW/m$^2$K which seems to suggest a size effect for this method; however, Chalopin \etal \cite{chalopin2012thermal} calculated similar conductances ($G\approx 0.8$ GW/m$^2$K) for Si/Ge super-lattice with spacing greater than about 10 nm. 
Furthermore, they reported up to an order of magnitude discrepancy with the results in \cref{landry2009effect} which used \eref{eq:GL} with a non-equilibrium method using a localized heat source and sink.
Also, given the consistency of the data in \fref{fig:tilt_conductivity} with the series resistor model, \eref{eq:GL}, it appears that all the systems are above the limit at which the majority of phonons travel ballistically from interface to interface which induces size effects in $G$ \cite{koh2009heat,giri2016heat}.

\begin{table}
\centering
%%%%%%%%%%%%%%%%%%%%%%%%%%%%%%%%%%%%%%%%%%%%%%%%%%%%%%%%%%%%%%%%%%%%%%%%
\begin{tabular} {| c | c | c | c |}
\hline
\multicolumn{4}{|c|}{tilt {[}nm$^{-1}${]}}\\
 0.0267  &   0.0535  &   0.1070  &   0.2140 \\
\hline
\hline
    28.15 $\pm$ 0.69  &      25.42 $\pm$ 0.60  &      20.71 $\pm$ 0.32  &      14.78 $\pm$ 0.39 \\
    28.68 $\pm$ 1.15  &      28.57 $\pm$ 0.92  &      26.09 $\pm$ 0.47  &      22.07 $\pm$ 0.60 \\
\hline
\end{tabular}
%%%%%%%%%%%%%%%%%%%%%%%%%%%%%%%%%%%%%%%%%%%%%%%%%%%%%%%%%%%%%%%%%%%%%%%%
\caption{Thermal conductivity (in W/m-K) as a function of interface defect (210 tilt) densities.
The upper entry in the table is the conductivity perpendicular to the planar defect and the lower entry is the conductivity parallel to the planar defect.
}\label{tab:kappa_interface}
\end{table}

\section{Conclusion} \label{sec:discussion}

In this work we examined the effects of defects and their interactions on the thermal conductivity of LiF using GK techniques and created models from these sample with embedded uncertainty estimates.
The uncertainty estimates were important since the model was used to predict the thermal conductivity of a sample that was rapidly compressed and contained a network of dislocations and point defects.
The accuracy of the model confounded by the expense of obtaining sufficient GK samples to drive the statistical noise to zero and the complexities of characterizing all the aspects of the defect structure that influence thermal conductivity.
Dislocations and other defects appear at relatively low uniaxial compression of LiF and the dislocation structure is complex and entangled, unlike the computational cells we used to calibrate the model.
Determining how to efficiently sample the effects of complex spatial correlations of dislocations on thermal conductivity is left for future work.

If we consider the defect types we studied representative of those created in LiF in high rate experiments, it is apparent that reductions on the order of a factor of two in the thermal conductivity of LiF can occur at very high defect densities.
We did not observe larger reductions in thermal conductivity apparently since the crystal structure in these highly defected system is still largely intact and phonons propagation is relatively unperturbed.
One basic type of defect, a volume defect such as a void, was not studied but may have significant effects on thermal conductivity if present in sufficient density \cite{zhou2012effects}.
We did consider voids, which are essentially clusters of point vacancies which have a tendency to aggregate\cite{catlow1981defect,lee2011effects}, but they were not stable at the pressure we selected and with the MD model we used.
Also, in contrast to the purely mechanical loading we employed to create a representative system, recent results by Zhou \etal \cite{zhou2016molecular} demonstrate that dislocation motion in an electric field can leave a  significant number of vacancies in its wake which is relevant to the use of LiF in pulsed power experiments.

In conclusion, the methodology of creating a model of the dependence of thermal conductivity on defects we developed successfully represented the data and its errors, is extensible to a wider range of defects, and is transferable to other materials.
Furthermore, it exposed trends in the dependence on defects not seen in the available approximate analytical models.

\section*{Acknowledgements}
We appreciate the use of LAMMPS \cite{lammps} and OVITO \cite{ovito}.
Sandia National Laboratories is a multimission laboratory managed and operated by National Technology and Engineering Solutions of Sandia, LLC., a wholly owned subsidiary of Honeywell International, Inc., for the U.S. Department of Energy?s National Nuclear Security Administration under contract DE-NA-0003525.


\end{document}